\documentclass[journal=jpccck,manuscript=article]{achemso}
\usepackage{graphicx}
\usepackage{epsf}
\usepackage{ifthen}
\usepackage{amsmath}
\usepackage{amssymb}
\usepackage{subfigure}
\usepackage{psfrag}
\usepackage{float}
\usepackage{subeqnarray}
\author{Sandeep Kumar Singh}
\affiliation{Universiteit Antwerpen, Department of Physics,
Groenenborgerlaan 171, BE-2020 Antwerpen, Belgium.}
\author{S. Costamagna}
\affiliation{Facultad de
Ciencias Exactas Ingenier{\'\i}a y Agrimensura, Universidad Nacional
de Rosario and Instituto de F\'{\i}sica Rosario, Bv. 27 de Febrero
210 bis, 2000 Rosario, Argentina.}
\author{M. Neek-Amal}
\email{mehdi.neekamal@gmail.com}
\affiliation{Universiteit Antwerpen, Department of Physics,
Groenenborgerlaan 171, BE-2020 Antwerpen, Belgium.}
\author{F. M. Peeters}
\affiliation{Universiteit Antwerpen, Department of Physics,
Groenenborgerlaan 171, BE-2020 Antwerpen, Belgium.}
\title{Melting of Partially Fluorinated Graphene: From Detachment of Fluorine Atoms to Large Defects and Random Coils}
\begin{document}
\begin{abstract}
The melting of fluorographene is very unusual and depends strongly
on the degree of fluorination. For temperatures below 1000\,K, fully
fluorinated graphene (FFG) is thermo-mechanically more stable than
graphene but at $T_m\approx2800$~K FFG transits to random coils
which is almost twice lower than the melting temperature of
graphene, i.e. 5300~K. For fluorinated graphene (PFG) up to 30$\%$
ripples causes detachment of individual F-atoms around 2000~K while
for $40$-$60\%$
 fluorination, large defects are formed beyond ~1500~K and beyond 60$\%$ of fluorination F-atoms remain bonded to graphene until melting.
 The results agree with recent experiments on the dependence of the reversibility of the fluorination process on the percentage of fluorination.
\end{abstract}

\section{Introduction}
 Several distinct atomic arrangements of adatoms (fluor,
hydrogen, chlorine, etc) have been proposed for tuning the electronic
properties of graphene (GE)~\cite{Nair,zob,elias}. The main
advantage of using fluor is that C-F bonds are energetically more
stable than e.g. C-H ones~\cite{sofo,leenaers1} because F-atoms posses
larger binding and desorption energies to C than
H-atoms~\cite{Nair}. Furthermore, fluorination is easier to control
via temperature and by reactant gases leading to reproducibly precise
C/F stoichiometries~\cite{withers1,Lee1,Robinson}. In the presence of F adatoms C-bonds in graphene transit from  {\it
sp$^2$} to  {\it sp$^3$} hybridization, which turn the conjugated,
graphitic C-C bonds into single C-C bonds. The lattice structure
results in an Angstrom scale out-of-plane buckled shaped membrane
known as chair configuration~\cite{sofo}  that
influences the high temperature stability of FG. Owing to the
properties mentioned above, a complete understanding of
the thermal behavior of the FG sheet is hence very important. 

Previous studies have shown that, different from
graphene, fully fluorinated  and also
hydrogenated~\cite{Sandeep,Costamagna} graphene, are more rigid for
temperatures up to $1000$~K. The situation beyond this temperature and up to
melting is not yet fully understood. Raman spectroscopy experiments
revealed that the fluorination time and the $N_C/N_F$ ratio (where $N_C$ and $N_F$ are
the number of carbon and fluor atoms, respectively)
are two important key factors in the preparation of FG~\cite{Nair,zob}.
It was found that the process can be reversed for low coverage and during
the fluorination process large membrane holes could appear due to
losses of C atoms even at room temperature~\cite{vanDuin,cheng}. Despite
this macroscopic information, the microscopical features and
temperature stability is  not understood. The aim of this letter is to
provide such a detailed microscopic understanding for different level of F coverage
and to explain recent experiments.
 We investigate the importance of the $N_C/N_F$ ratio and  the modifications induced by vacancy defects
 on the stability of FG and the melting process.

\section{Simulation Method}
 Molecular
dynamics (MD) simulations using reactive force fields
(ReaxFF~\cite{van, Duin, Chenoweth}) present in the large-scale
atomic/molecular massively parallel simulator (LAMMPS)
code~\cite{Plimpton} are used. ReaxFF is a bond-order-dependent potential that describes bond formation and dissociation. Many body interactions such as the valence angle and torsional interactions are formulated as function of bond order. Non bonded interactions, e.g. Coulomb and van der Waals interactions, are included for all pair atoms which are not well treated usually  by quantum mechanical methods such as density functional theory (DFT)~\cite{Boukhvalov2009}. Excessively close range interactions are avoided by shielding. To account for the van der Waals interaction, a distance-corrected Morse-potential is used. ReaxFF uses the  geometry dependent charge calculation scheme (EEM scheme) of Mortier~\textit{et al}~\cite{mortier}. Intra-atomic contributions of atomic charges which is required to polarize the atoms, are included in the energy scheme, which allows to apply the force fields  to ionic compounds~\cite{Touhara2000}.  The system energy in ReaxFF consists of a sum of terms:
\begin{eqnarray}\nonumber
E_{sys} = E_{bond} + E_{under} + E_{over} + E_{lp} + E_{val} + E_{pen} +
&&\\ \nonumber
E_{tors} + E_{conj} + E_{vdWaals} + E_{Coulomb}.
\end{eqnarray}
A detailed general description of each of these terms and their functional forms can be found in the original work~\cite{van} and for fluorinated graphene in our previous paper~\cite{Sandeep}.  ReaxFF is able to predict very precisely the equilibrium C-F bond length and the C-F bond dissociation energy, close to  DFT based energies and geometries for a number of molecules and reactions.
  We considered square
shaped computational unit cells having $N_C$=1008 C-atoms in the graphene
sheet with partial fluorination up to fully fluorinated
graphene (FFG). The simulations were performed in  the NPT (P=0) ensemble with periodic
boundary conditions.  Temperature was maintained by the
Nos\'{e}-Hoover thermostat~\cite{Hoover} and the MD time-step was
taken to be 0.1~fs.
\begin{figure}[t]
\includegraphics[width=0.80\textwidth]{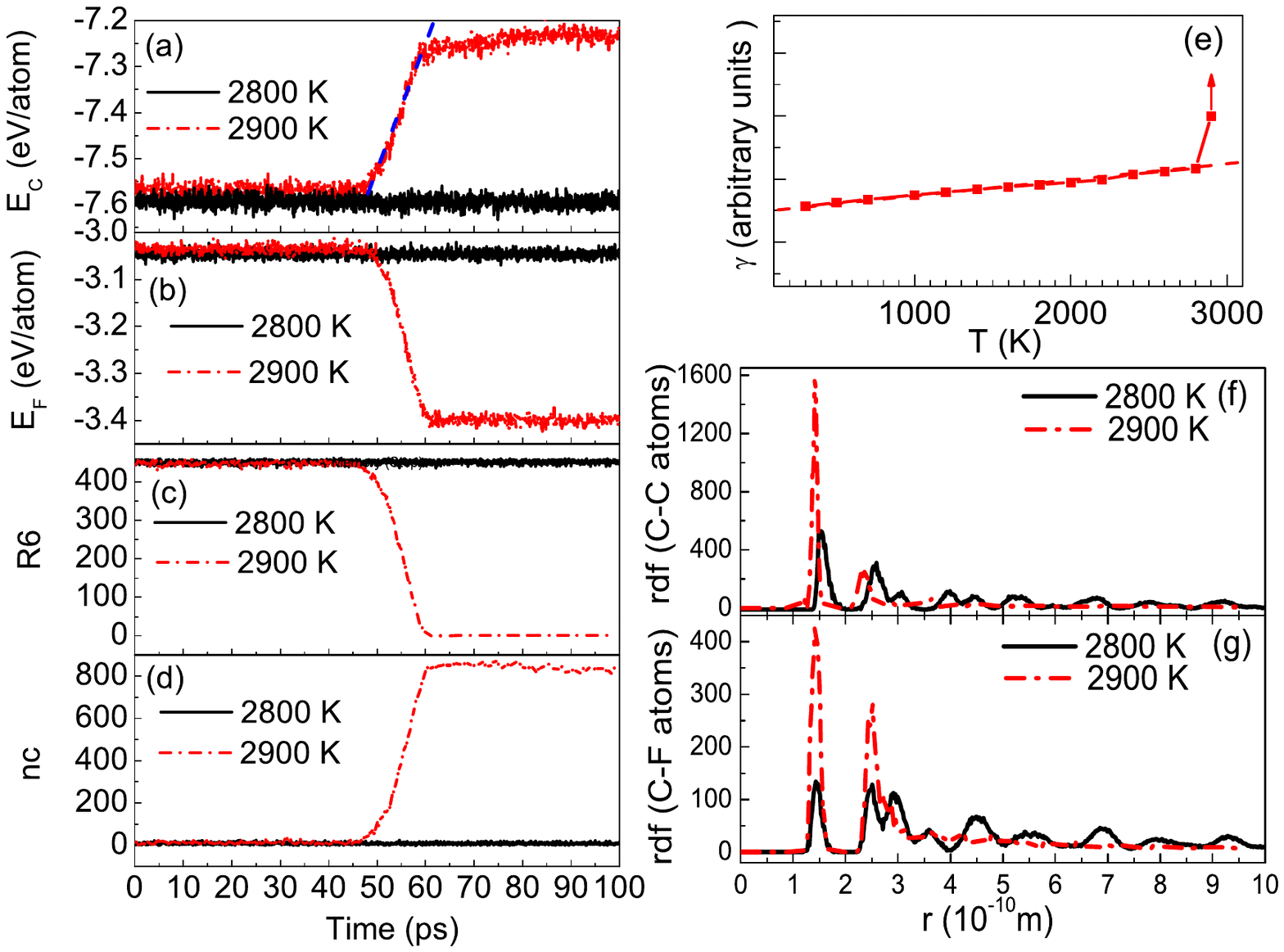}
\vspace{-0.6cm} \caption{(Color online) Fully fluorinated graphene. (a) $E_C$, (b)
$E_F$, (c) number of six-membered rings R6, and (d) chains (nc) with
more than three connected twofold-coordinated atoms, as a function
of time for T=$2800$~K and at T=$2900$~K where
melting occurs between these temperature range. (e) Modified Lindemann parameter $\gamma$ versus
temperature. (f) C-C and (g) C-F radial distribution function.
}\label{defect}
\end{figure}
The partial covered samples (PFG) were designed by adding randomly F
atoms equally distributed on both sides of an initially flat
graphene lattice~\cite{Sandeep} such that each carbon gets maximum one fluorine atom.  The dependence of the averaged
lattice parameter with increasing F concentration was found to be in
agreement with recent results~\cite{partial-ff}.

To account for the melting transition we analyzed the variation of
the total potential energy $E_T$ per atom with temperature
identifying partial contributions from C-atoms ($E_C$) and F-atoms
($E_F$). The Lindemann criterion was used to characterize the
ordered state by considering the modified
parameter $\gamma$, used previously for 2D systems~\cite{Bedanov,
Zheng1}.

\begin{figure*}[t]
\centering
\begin{center}
\includegraphics[width=0.30\textwidth]{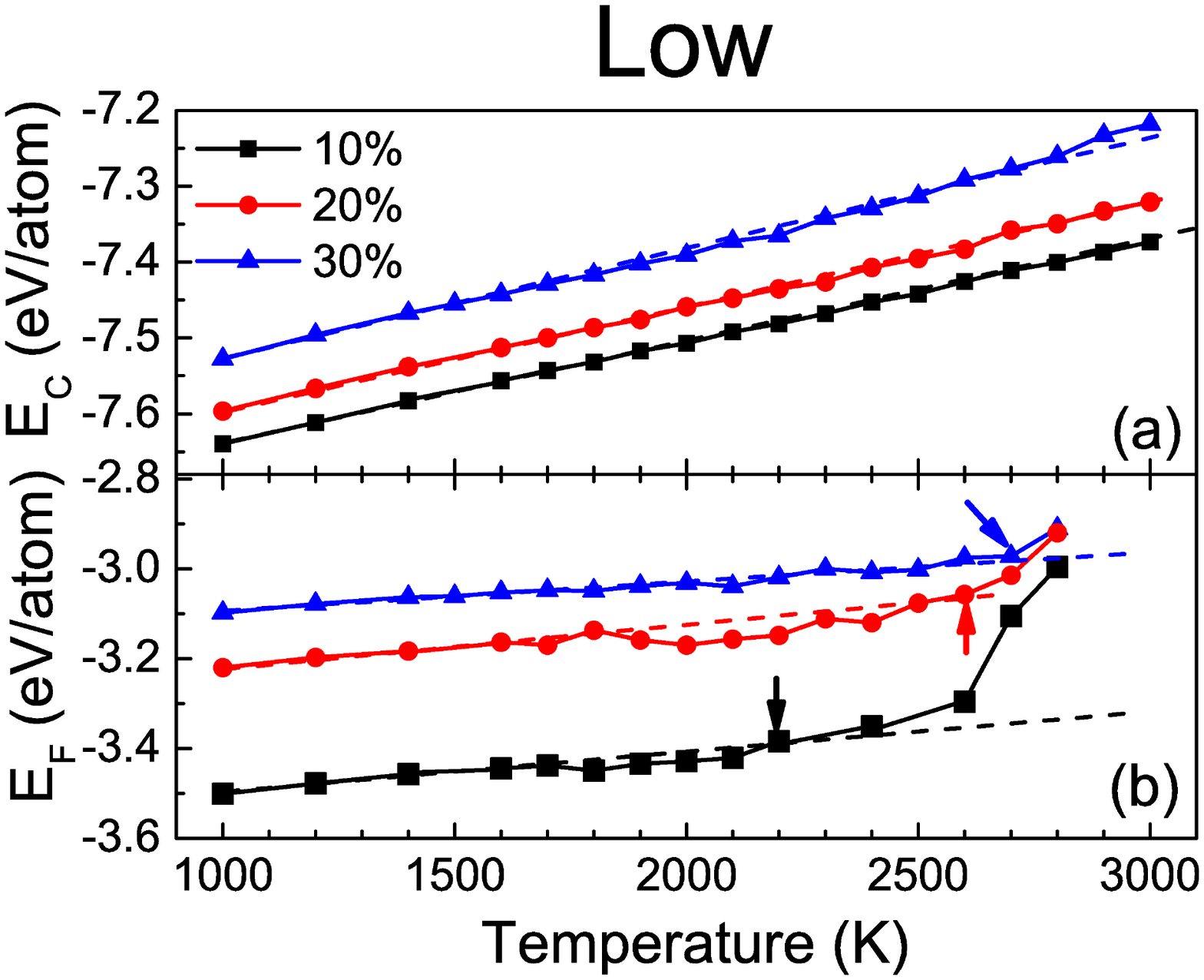}
\hspace{0.20cm}
\includegraphics[width=0.30\textwidth]{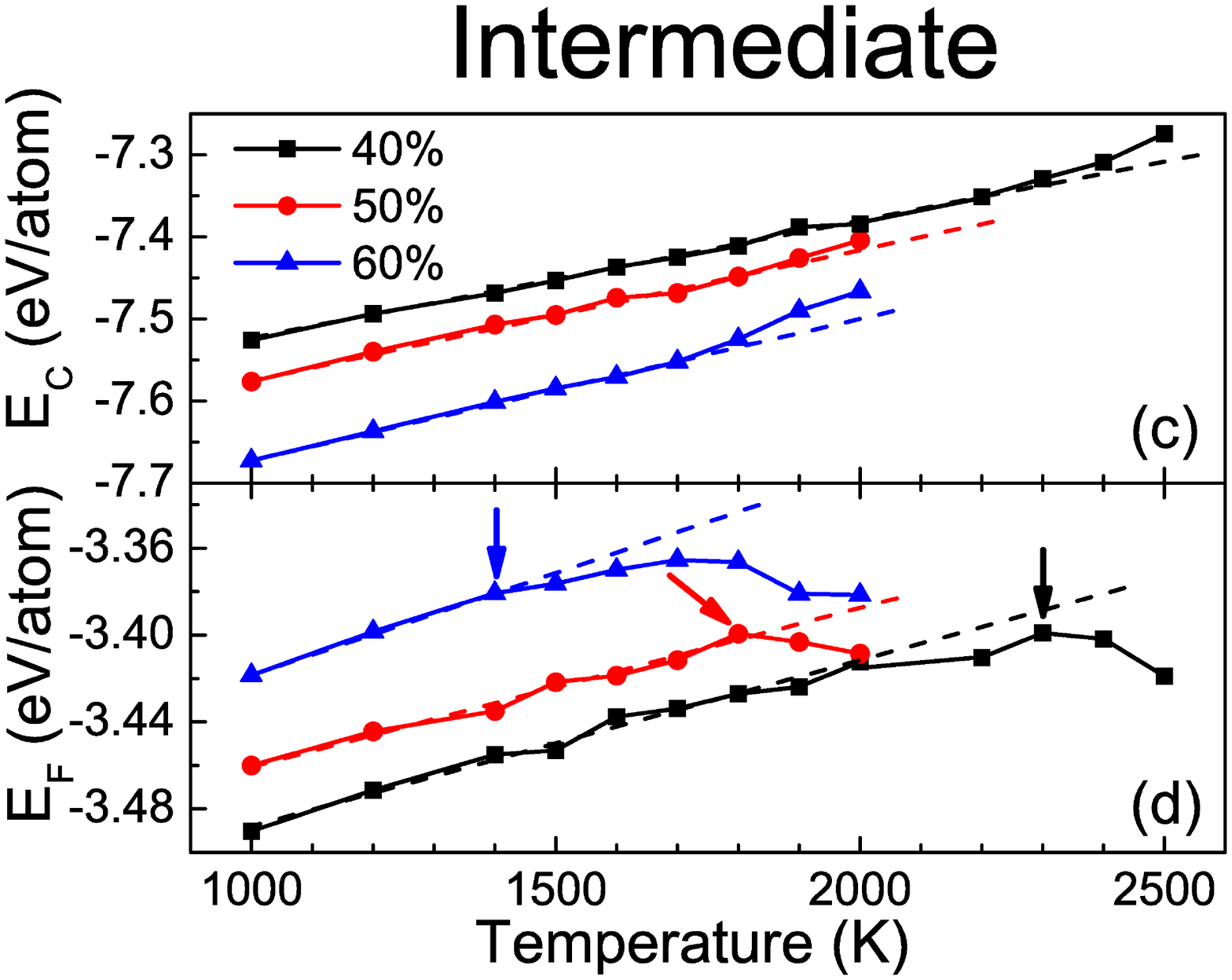}
\hspace{0.20cm}
\includegraphics[width=0.30\textwidth]{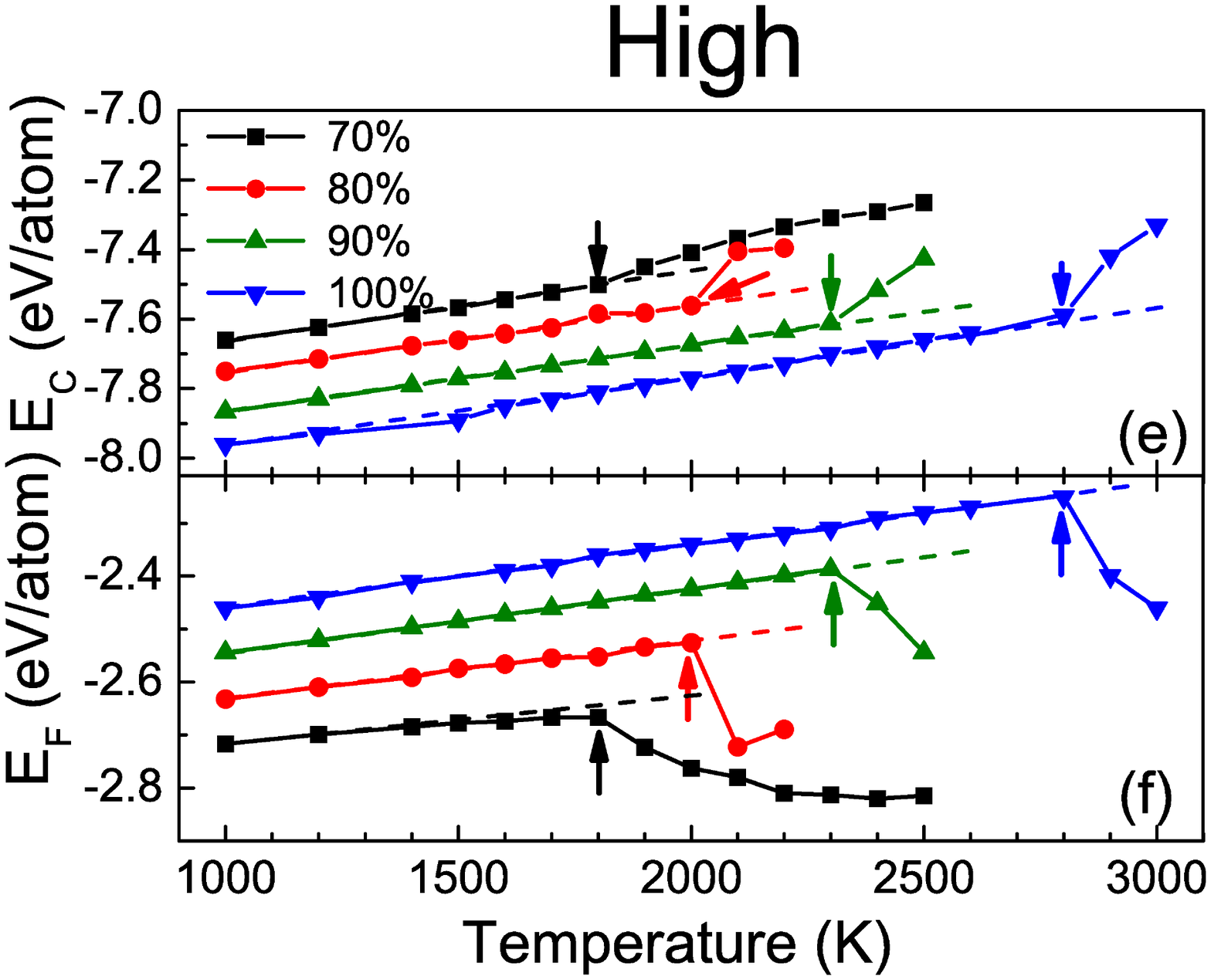}
\end{center}
\vspace{-0.6cm}\caption{(Color online) Dependence of $E_C$ (upper panels) and $E_F$
(lower panels) versus temperature for the three regimes of melting
indicated. Curves were shifted for a better
comparison ((i) for $E_C$: 30, 40 and 60$\%$ fluorination, curves are shifted by 0.05, 0.45 and -0.5 eV/atom, respectively. (ii) for $E_F$: for 20, 30, 70, 80, 90, and 100$\%$ fluorination, curves are shifted by 0.3, 0.4, 0.65, 0.7, 0.75, and 0.8 eV/atom, respectively). Notice that  $E_F$ increases in (b) while it decreases in (d) and (f) which is an indication of
the evaporation of fluor in ``low" concentration regime and ring/defect formation and their saturation by F in
``intermediate" and ``high" concentration regimes". }\label{pot}
\end{figure*}


\section{Results} We first analyze the case of FFG.
Figures~\ref{defect}(a,b) show the variation of the potential energy
per atom of carbon and fluor atoms, i.e. $E_C$ and $E_F$
respectively, with time at $2800$~K and $2900$~K. The sharp increase
(decrease) in $E_C$, which is about 4.5$\%$ ($E_F$ about $10\%$), is
a signature of melting at $2900$~K. During melting, (10\,ps) the
number of six-membered rings (R6) of the crystalline phase is
reduced (Fig.~\ref{defect}(c)) and chains composed by single C-atoms
bonded to F-atoms are formed (Fig.~\ref{defect}(d)). The  melting
temperature $T_m=2800$~K is further confirmed by the Lindemann
parameter $\gamma$ (Fig.~\ref{defect}(e)). Due to the strong
covalent nearest-neighbor C-C interaction $\gamma$ increases
linearly up to close the melting temperature where it diverges.

After melting the C-atoms in the single chains remain bonded to the
F-atoms, i.e. a spaghetti of 3D-polymers constructed from C-F
monomers (a snap shot of the molten FFG will be shown in
Fig.~\ref{phase}(c)).  For large simulation time the molten
structure is composed of C-chains which form an entangled
three-dimensional network which looks more like a polymer gel than a
simple liquid~\cite{Sandeep2}. The larger reduction in $E_F$
suggests that  F atoms prefer to be bonded to  carbon atoms of
rings/chains rather than the carbons of GE. This indicates that in
experiments during fluorination  F atoms first prefer to be bonded
to defective regions, where the environment  is more similar to a
C-chain. The radial distribution function (rdf) indicates that the
C-C distance in chains is shorter and after melting (double bonds
appear) only one significant peak remains (see
Fig.~\ref{defect}(e)). However the C-F rdf in Fig.~\ref{defect}(e)
shows that after melting there are two significant peaks which
correspond to the  appearance of -C-F$_2$ and -C-F bonds.
In the case of PFG, conjugated C-C
double bonds in the non-fluorinated parts coexist at low
temperatures with covalent C-F bonds in
corrugated fluorocarbon regions~\cite{Zhang11, Giraudet}. However the behavior  at
higher temperature is more complicated. We found
three different melting processes depending on the fluor percentage:

\subsection{Low fluorination} ($10$-$30\% $). In Figs.~\ref{pot} (a,b) we display
$E_C$ and $E_F$ against temperature. The main feature here
 is  evaporation  of F-atoms and the formation of large rings, e.g. R10, R14. While $E_C$ increases linearly
against temperature due to a smooth expansion of the C-C
inter-atomic distance (small deviations are due to the formation of
large rings), because of C-F bond breaking, $E_F$ instead behaves
non-linear (and obtains higher energies: F atoms are in the gas
phase, see Fig.~\ref{phase}(a)) with Temperature (indicated by the
arrows). Detachment starts around $2000$~K and continues until
almost all the F-atoms are removed from the
membrane~\cite{big-rings}. With further increase in temperature, the
Lindemann criterion indicates that melting occurs at 5100\,K,
4500\,K and 3500\,K for 10, 20 and 30$\%$ fluorination, respectively
and F-atoms are less likely to be bonded to the GE. \emph{Thus, the
linear increase in $E_C$, Fig.~\ref{pot}(a), shows that the PFGs
with $N_F/N_C\leq30\%$ turns into pristine graphene (with a few
number of rings)} around 3000 K.

\subsection{Intermediate fluorination} ($40$-$60\% $). In this regime
 at about $T\approx 1500$~K, $E_C$
shows small deviations from the linear behavior due to the formation of
large defective rings (Fig.~\ref{pot}(c)). In contrast to the case
of low $N_F/N_C$ ratio, $E_F$ shown in Fig.~\ref{pot}(d), becomes
more negative since  F-atoms instead of being detached from the
sheet, now break the original local bonds and are transferred to the
edges of the rings (a snapshot of this configuration is shown
in Fig.~\ref{phase}(b)). Then, at the  melting temperature clusters of C-F are formed near
the defects and they detach from the sheet. The microscopic
configurations consist  of a mixture of  2D and 3D phases.

\subsection{ High fluorination} ($70$-$90\% $). Finally, in the
high-fluorination regime with increasing temperature large ripples
develop while F atoms remain bonded to graphene until melting
occurs. Now, $E_C$ receives positive contributions due to the
breaking of C-C bonds at melting (Fig.~\ref{pot}(e)) and since the
C-F bond becomes stronger $E_F$ is reduced (Fig.~\ref{pot}(f)). The
microscopical features of melting in this case posses similar
characteristics as FFG. Notice that arrows in Fig.~\ref{pot} are not
pointing  necessarily  to the melting transitions, instead they
indicate mostly evaporation (b) and defect formation thresholds
Figs.~\ref{pot}(d,f). \emph{The reduction in  $E_F$ indicates the
appearance of polymers consisting of C-F monomers.} The average
melting temperatures $T_m$ were estimated from the behavior of the
Lindemann parameter. Starting from $60\%$, when we increase further
the F-content $T_m$ is seen to pass a minimum around $70\%$ and then
increases due to the suppression of long wave-length
ripples~\cite{Sandeep}.

\begin{figure}
\includegraphics[width=0.80\textwidth]{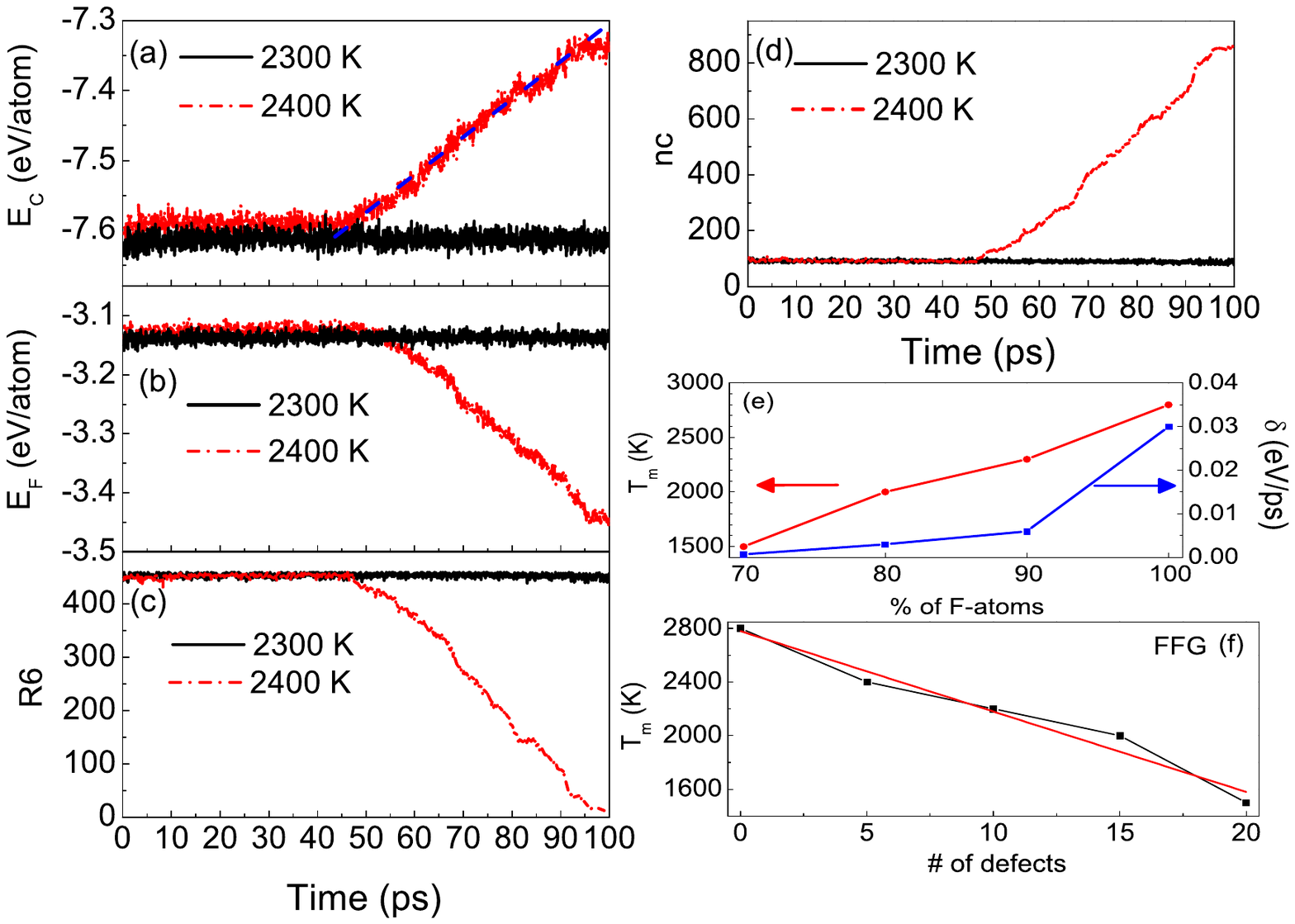}
\vspace{-0.6cm}\caption{(Color online) Partially fluorinated graphene $90\%$. (a)
Total energy of C-atoms, (b) total energy of F-atoms, (c) number of
six-membered rings R6, (d) number of chains (nc) with more than
three connected twofold-coordinated atoms, as a function of time at
T=$2300$~K (below melting) and at T=$2400$~K where melting occurs,
(e) $\delta$ and T$_m$ versus percentage of F atoms, and (f)
variation of the melting temperature $T_{m}$ against the number of
defects in fully fluorinated graphene. }\label{90}
\end{figure}

We depict the energy variation during melting for a typical case of $90\%$ fluorination in  Figs.~\ref{90}(a,b). Here, the number of hexagons
approach zero and many single chains are formed as shown in Figs.~\ref{90}(c,d).
From the slope of $E_C$ versus time we obtain the rate at which the
hexagonal crystal structure  of the sample becomes lost  (e.g.
dashed blue curve in Fig.~\ref{90}(a) shows $\delta=dE_C/dt$). Values of $\delta$ are
shown in Fig.~\ref{90}(e) together with $T_m$. The main result is that
the larger the ratio $N_F/N_C$, the  higher the melting temperature
and the faster that melting takes place.

\section{ Effect of vacancies on the melting of FFG} It is important to
study the effect of the presence of atomic vacancies in FFG on the
melting temperature. We performed several additional simulations for
FFG with $p$ number of vacancies which were randomly distributed
over FFG, i.e. $N_C\rightarrow N_C-p$ and $N_F\rightarrow N_F-p$.
The presence of atomic defects in the GE sheet makes it less
stiff~\cite{prb82neek} and consequently results in a lowering of the
melting temperature, see Fig.~\ref{90}(f). The melting temperature
is fitted by the red line T$_m$(N)=a+bN, where a=(2780$\pm$75)~K,
b=(-60$\pm$6)~K and N is the number of vacancies. These results
clearly indicate that T$_m$ decreases sharply as the number of
defects in the FFG increases.

\section{Discussion} FFG is an insulator with a large band gap while
graphene shows metallic properties, thus one naturally expects the
transition from GE to FFG by increasing the ratio $N_F/N_C$. This
gives the appropriate melting trend for a typical metal to an
insulator. However, the melting of GE and FFG is different because
of their 2D-nature. The existence of flexural phonons makes GE and
FFG also different from typical 3D crystals~\cite{leenaers1} and the
bending rigidity is a temperature dependence parameter~\cite{seba,
sandeep_BN}. Although at low temperature FFG is  thermo-mechanically
more stable than GE, by increasing temperature, due to the
excitation of the vibrational C-F mode, in FFG the entropy term
increases faster with T which is being responsible for the observed
lower melting temperature.

\begin{figure}
\vspace{-0.6cm}
\includegraphics[width=0.80\textwidth]{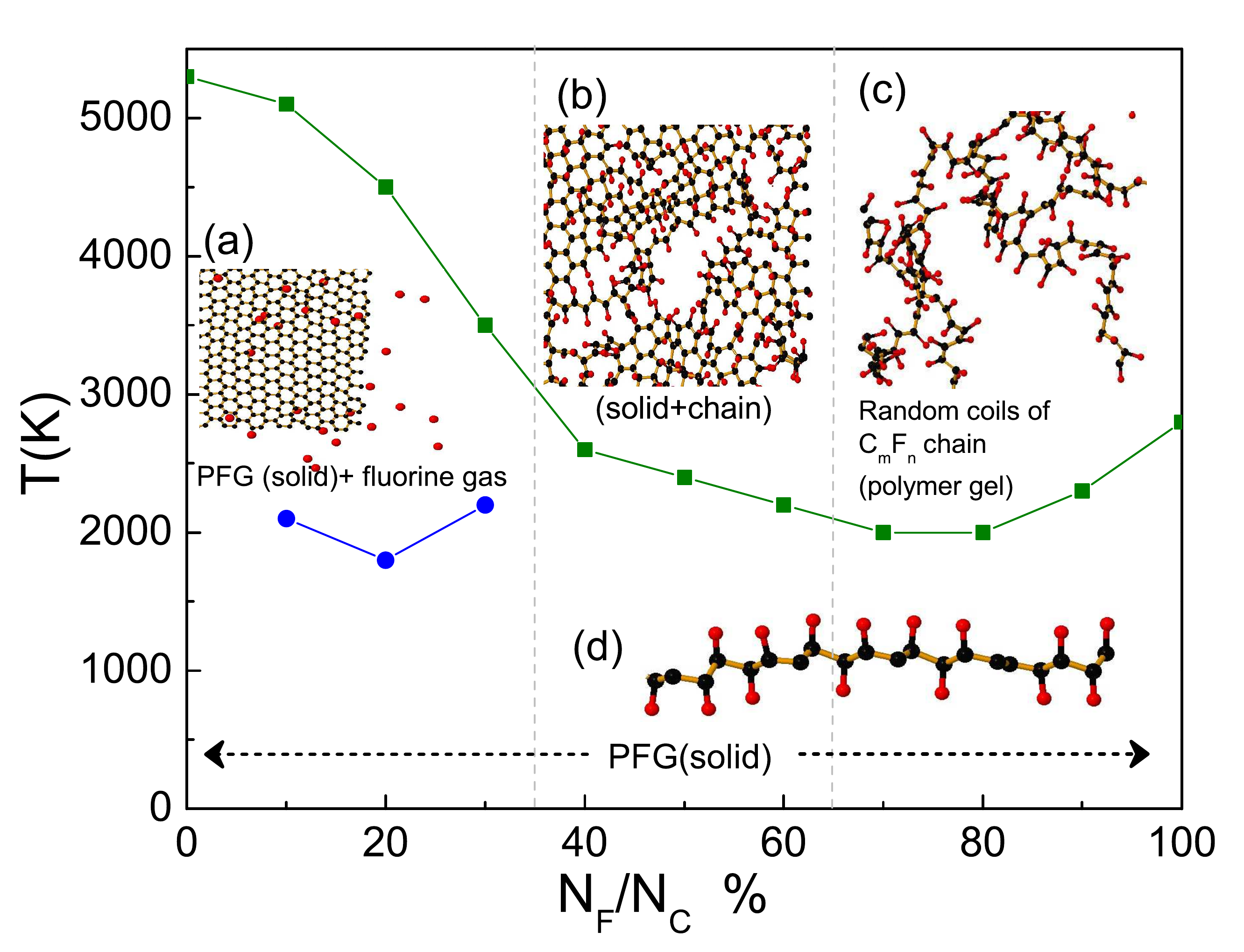}
\caption{(Color online) Melting phase diagram for fluorinated graphene. Circular
symbols refer to the evaporation of F atoms (blue circles). The
insets show the top view of the simulated FG before (a) and after
melting (b,c). The inset (d) is a side view of the simulated PFG
with $N_F/N_C=80\%$.  }\label{phase}
\end{figure}

In PFG the distribution of masses through the system is
non-uniform, hence the vibrational frequencies are not well
defined and are position dependent. This randomness in the system
produces very large out-of-plane fluctuations even at low temperature.
 This  broadening in the frequency range brings the system closer to the melting
transition point.

Moreover, for a low ratio $N_F/N_C$ when we heat the system F-atoms
are evaporated in order to reduce the total energy. Then, the system
behaves like pristine graphene and evaporated F atoms have no chance
to be re-bonded to the system. For intermediate ratio $N_F/N_C$, the
concentration of F atoms in some random domains make the system
unstable due to the growth in the mean square value of the height
fluctuations $\langle h^2\rangle$ resulting in the formation of
ring-defects with increasing temperature and the melting of this new
defected FG is more complex. Finally, for high fluorination we
deduce that the melting temperature is proportional to the coverage
percentage, e.g. decreasing the coverage percentage from 100$\%$ to
90$\%$ (80$\%$) decreases the melting temperature with about 16$\%$
(30$\%$).

Lets now compare our results  with recents
experiments~\cite{Nair,zob}. Raman spectra of graphene after
exposure to atomic F shows dramatic changes induced by fluorination:
the D peak emerges at early times of fluorination and the 2D peak is
suppressed which indicates that F atoms are bonded to C atoms and
atomic defects appear. The process of fluorination has been shown to
be reversible only when the exposure time is relatively short
($<20$~hours) and the concentration of F atoms is presumably low.
FFG is obtained only for longer exposure time, beyond 30 hours,
where the evaporation of F atoms and the restoring to pristine
graphene is impossible. In our simulations, we found that for low
coverage percentage (less than 40$\%$) increasing the temperature
causes evaporation of F atoms and almost pristine graphene can be
recovered in agreement with experiment. However, percentage beyond
40$\%$ are related to long time fluorination ($>$ 20 hours) where
the process was reported to be irreversible. The presence of
vacancies decrease dramatically the melting temperature. Indeed, in
experimental PFG samples containing a few vacancies it was found
that when they were heated up to much lower temperatures
($\sim$500-700~K) it was possible to restore to not-perfect pristine
graphene~\cite{Nair}.

\section{conclusion}
The phase diagram displayed in Fig.~\ref{phase}  summarizes our
results: green squares refer to the transition  temperature and
blue circles indicate the starting point of evaporation of F-atoms. Two
insets (b) and (c) show a top views of the system after the transition points, inset
(a) shows top view of a snap shot of a system which
lost fluor and inset (d) shows a side view of a typical PFG at low temperature.
The PFG above the blue symbols for $N_F/N_C<40\%$ looses fluor and
become pristine graphene with some defects and above the green line it
transits to the liquid phase, however for
$N_F/N_C\sim$40-60$\%$ above the green line and below ($\sim$5000\,K)
PFG does not loose  F and very slowly transits to the  liquid
phase so that below 5000\,K the liquid and the solid phases  coexist.
For larger $N_F/N_C\sim$60-100$\%$ above the green curve PFGs rapidly
transits to a 3D liquid phase. The minimum melting
temperature is found to be around 70$\%$ fluorination.
The vertical dashed lines
separate three different regions for different percentage of F
atoms.  Our findings are therefore consistent with the experimental
reversibility of the fluorination process in single layer graphene.

\begin{acknowledgement}
This work was
supported by the EU-Marie Curie IIF postdoc Fellowship/299855 (for
M.N.-A.), the ESF-Eurographene project CONGRAN, and the Flemish
Science Foundation (FWO-Vl). Financial support from   the
Collaborative program  MINCyT(Argentina)-FWO(Belgium) is also
acknowledged.
\end{acknowledgement}


\begin{thebibliography}{13}
\bibitem{big-rings} The small reduction in the potential energy curves $E_F$
in the temperature range $1400-1700$~K is due to the appearance of a
few large defects (i.e. big rings R10 and R14).
\end{thebibliography}
\end{document}